\pdfoutput=1
\documentclass[%
 reprint,
 superscriptaddress,
 amsmath,amssymb,
 floatfix,
]{revtex4-2}

\usepackage[utf8]{inputenc}
\usepackage[T1]{fontenc}

\usepackage{graphicx}% Include figure files
\usepackage{xcolor}
\usepackage{dcolumn}% Align table columns on decimal point
\usepackage{bm}% bold math
\usepackage{hyperref}  % add hypertext capabilities
\hypersetup{colorlinks,allcolors=blue}
\usepackage[all]{hypcap}
\usepackage[retain-explicit-plus,range-phrase=\textendash,range-units=single]{siunitx}
\usepackage{etoolbox}
\usepackage{booktabs}
\usepackage{multirow}
\newcolumntype{P}[1]{>{\centering\arraybackslash}p{#1}}

\usepackage{array}
\newcolumntype{H}{>{\setbox0=\hbox\bgroup}c<{\egroup}@{}}

\makeatletter
\renewcommand{\p@section}{}
\renewcommand{\p@subsection}{}
\renewcommand{\p@subsubsection}{}
\renewcommand{\figurename}{Figure}
\renewcommand{\tablename}{Table}
\renewcommand{\fnum@figure}{\textbf{\figurename~\thefigure}}
\renewcommand{\fnum@table}{\textbf{\tablename~\thetable}}
\makeatother

\renewcommand{\thetable}{\arabic{table}}

\usepackage[capitalize,noabbrev,nameinlink]{cleveref}
\crefformat{equation}{#2eq~#1#3}  % remove parantheses for eq
\crefmultiformat{equation}%
{\edef\crefstripprefixinfo{#1}#2eqs~#1#3}%
{ and~#2\crefstripprefix{\crefstripprefixinfo}{#1}#3}%
{, #2\crefstripprefix{\crefstripprefixinfo}{#1}#3}%
{, and~#2\crefstripprefix{\crefstripprefixinfo}{#1}#3}

\begin{document}

\title{Comprehensive understanding of H adsorption on MoO\texorpdfstring{\textsubscript{3}}{3} from\texorpdfstring{\\}{ }systematic \textit{ab initio} simulations}

\author{Yuji Ikeda}
\email{yuji.ikeda@imw.uni-stuttgart.de}
\affiliation{Institute for Materials Science, University of Stuttgart, Pfaffenwaldring 55, 70569 Stuttgart, Germany}

\author{Deven Estes}
\affiliation{Institute of Technical Chemistry, University of Stuttgart, Pfaffenwaldring 55, 70569, Stuttgart, Germany}

\author{Blazej Grabowski}
\affiliation{Institute for Materials Science, University of Stuttgart, Pfaffenwaldring 55, 70569 Stuttgart, Germany}

% \date{\today}

\begin{abstract}
During many of its applications (especially as a catalyst support material), MoO\textsubscript{3} acts as a medium for hydrogen storage via hydrogen spillover (H atom donation from proton and electron sources to a support), for which the energetics of H atoms on MoO\textsubscript{3} are of importance.
Despite the seeming simplicity of hydrogen spillover, previously reported \textit{ab initio} results for the H adsorption on MoO$_3$ contradict both experimental work and other \textit{ab initio} results. In the present study, we resolve these discrepancies and provide a comprehensive \textit{ab initio} understanding of H adsorption for MoO\textsubscript{3} in the bulk and on the surface. To this end, we systematically investigate various exchange--correlation functionals and various H concentrations, and we carefully track the various relevant H positions.
For a dilute H concentration, the asymmetric oxygen site (O\textsubscript{a}) is found to be energetically the most favorable.
With increasing H content, the difference of the H adsorption energies between the terminal (O\textsubscript{t}) and the O\textsubscript{a} sites becomes smaller.
Previous contradictions are ascribed mostly to the disregard of the H position along the O\textsubscript{a}--O\textsubscript{a} zig-zag chains in the intrabilayer region.
Using the modern non-empirical strongly-constrained and appropriately-normed (SCAN) meta-generalized gradient approximation (GGA), the dilute-limit H adsorption energies are obtained as \SI{-2.89}{\electronvolt/(H~atom)} and \SI{-2.97}{\electronvolt/(H~atom)} in the bulk and on the surface, respectively, and the activation energy of H diffusion between the O\textsubscript{a} sites as \SIrange{0.11}{0.15}{\electronvolt/(H~atom)}, consistent with previous experiments.
\end{abstract}

\maketitle

\section{Introduction\label{sec:introduction}}

\subsection{Motivation\label{sec:motivation}}

% \textcolor{C0}{[Global overview of MoO$_3$ or ($\alpha$-MoO\textsubscript{3})]}

The $\alpha$ phase of MoO\textsubscript{3} is a layered transition-metal oxide and a semiconductor with a wide band gap of \SI{2.8}{\electronvolt} as a single crystal~\cite{Scanlon_JPCC_2010_Theoretical}.
% We have to be careful that, many studies report the "thin-film" band gap, which is about 3.1--3.4 eV.
% The exp. band gap also depends on temperature.
% \cite{Carcia_TSF_1987_Synthesis}: thin film 3.1 eV
% \cite{Sabhapathi_ML_1994_Structural}: thin film: 3.24 eV at RT
% \cite{Sabhapathi_JMSL_1995_Growth}: thin film: 3.2 eV at RT
% \cite{Simchi_JAP_2013_Characterization}: thin film: 3.1--3.4 eV at 400 or 500 K
This oxide has a number of technologically important applications, e.g., as
a cathode material in rechargeable lithium ion batteries~\cite{Tsumura_SSI_1997_Lithium,Li_JPCB_2006_Vapor},
as a catalyst~\cite{%
Zeng_IC_1998_Chemical,
Uchijima_PCCP_2000_Catalytic,
Prasomsri_EES_2013_Effective,  % >200 citation
Ranga_CEJ_2018_Effect},
in photochromic and electrochromic devices~\cite{%
Yang_JPCB_1998_Microstructures,
Yao_N_1992_Photochromism},  % nature paper >500 citation
in optoelectronic nanodevices~\cite{Xiang_SR_2014_Gap},
as a gas sensor \cite{Comini_CPL_2005_Gas,  % MoO3 ``nanorod''
Ferroni_SABC_1998_MoO3},
and as a capacitive energy storage~\cite{Kim_NM_2016_Oxygen}.
%%%%%
% \cite{Haque_JMCA_2019_Ordered} non-metallic catalytic material for the hydrogen evolution reaction (HER).
% \textcolor{red}{Be careful that the crystal structure is different.}
%%%%%
% \textcolor{C0}{[H spillover]}
%%%%%
Among the various applications, MoO\textsubscript{3} is used as a support material for hydrogen storage~\cite{Uchijima_PCCP_2000_Catalytic} via the so-called hydrogen spillover mechanism~\cite{Li_JACS_2006_Hydrogen,Li_JPCB_2006_Hydrogen}.
An unusually high hydrogen capacity can be achieved 
via the dissociative chemisorption of H atoms on a nobel-metal catalyst and the subsequent migration of these H atoms onto the surface of a MoO\textsubscript{3} support material.
% \textcolor{C0}{[can refer to~\cite{Conner_CR_1995_Spillover}]}

% The activation energy for the H diffusion in the hydrogen bronze H\textsubscript{\textit{x}}MoO\textsubscript{3} is reported in some experimental reports~\cite{Slade_JSSC_1980_NMR,Ritter_BBPC_1982_Quasi,Ritter_JCP_1985_Structure}.

To understand the catalytic activity of $\alpha$-MoO\textsubscript{3} as well as the hydrogen spillover onto it, the energetics of H atoms are of critical importance, and in particular it is essential to know the energetically favorable H adsorption sites on MoO\textsubscript{3}.
In experiments, these issues have been addressed by considering hydrogen bronze, i.e., H$_x$MoO$_3$ with $0 < x \le 2$.
For a relatively low H content of $x \le 0.5$,
powder neutron diffraction~\cite{Schroeder_ZAAC_1977_Beitraege,Dickens_JSSC_1979_Elastic}
% Schroeder_ZAAC_1977_Beitraege: H0.5MoO3
% Dickens_JSSC_1979_Elastic: D0.36MoO3
and nuclear magnetic resonance~\cite{Ritter_BBPC_1982_Quasi,Ritter_JCP_1985_Structure}
% Ritter_BBPC_1982_Quasi: H0.35MoO3
% Ritter_JCP_1985_Structure: H0.2MoO3, H0.35MoO3
showed that H is bound to the asymmetric oxygen (O\textsubscript{a}) sites and resides in the intrabilayer region.
For higher H contents of $1.55 < x \le 2$, powder neutron diffraction~\cite{Dickens_SSI_1988_crystal,Anne_JPF_1988_Structure}
% Dickens_SSI_1988_crystal: D1.68MoO3
% Anne_JPF_1988_Structure: D1.65MoO3
showed that H is bound to the terminal oxygen (O\textsubscript{t}) sites and resides in the interbilayer region.
%%%%%
We can thus qualitatively expect that the H adsorption energy substantially depends on the O site as well as the H content.
It is, however, very difficult to address this dependence quantitatively solely by experiments.

% an analysis based on the power X-ray diffracition and the bond valence sum 
% \cite{Braida_CM_2005_Concerning} expected that

In principle, \textit{ab initio} calculations can help to provide a detailed understanding of H spillover onto MoO\textsubscript{3}.
H adsorption in bulk $\alpha$-MoO\textsubscript{3} was investigated in a few previous \textit{ab initio} studies~\cite{Braida_CM_2005_Concerning,Sha_JPCC_2009_Hydrogen}. H adsorption on the surface of $\alpha$-MoO\textsubscript{3} was likewise addressed in several papers 
in relation to the surface reaction of various adsorbates such as
the hydrogen atom itself~\cite{%
Chen_JCP_1998_density},
methyl radical~\cite{Chen_JCP_2000_density,Chen_JACS_2001_Chemical},  
ethylene~\cite{Yang_JPCC_2012_First},
acetaldehyde \cite{Mei_JPCC_2011_Density}
acetone~\cite{Shetty_JPCC_2017_Computational,Pan_IJHE_2019_first},
% Shetty: hydrodeoxygenation (HDO) from acetone to propylene
methanol~\cite{RellanPineiro_JPCL_2018_One},
and
formaldehyde~\cite{RellanPineiro_JPCL_2018_One}.
Most of these studies reported their computed H adsorption energies as well as the energetically favorable O sites for the H adsorption.
These results are, however, contradictory among each other (cf.~\cref{sec:review}) and sometimes even with respect to the experimental findings mentioned above.
These contradictions have not yet found any explanation, and this discrepancy has clearly impeded the establishment of reliable \textit{ab initio} results.

In the present study, we systematically investigate H adsorption for $\alpha$-MoO$_3$ in the bulk and on the surface based on \textit{ab initio} simulations.
We consider various H locations, even for the same O site (cf.~\cref{sec:H_sites}), as well as the impact of the H content.
We investigate various exchange--correlation functionals with and without consideration of the van der Waals (vdW) interaction.
We offer a comprehensive understanding and resolve the contradictions among the previously reported \textit{ab initio} results. Our results provide the basis for a better interpretation of experimental observations.

\subsection{Crystal structure of \texorpdfstring{$\bm\alpha$}{α}-MoO\texorpdfstring{\textsubscript{3}}{3}}

\begin{figure}
    \centering
    \includegraphics[width=\linewidth]{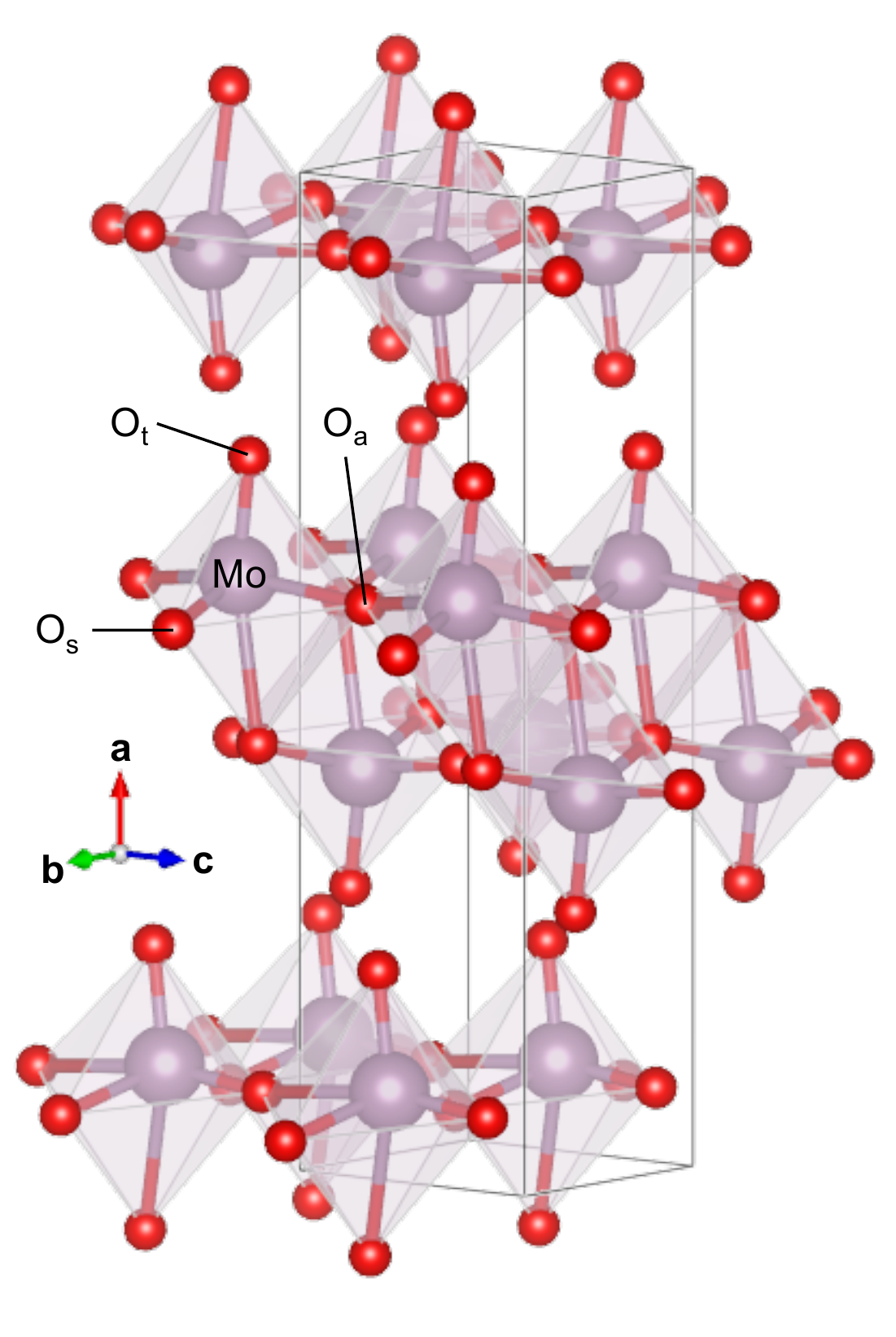}
    \caption{Crystal structure of $\alpha$-MoO$_3$.
    The purple and the red spheres represent Mo and O atoms, respectively.
    Visualization has been performed using VESTA~\cite{Momma_JAC_2011_VESTA3}.}
    \label{fig:structure_bulk}
\end{figure}

% \textcolor{C0}{[Description of the bulk $\alpha$-MoO$_3$ structure.]}

The ground state of MoO$_3$ is called the $\alpha$ phase (\cref{fig:structure_bulk}),
and it is characterized by the space group $Pnma$ (No.\,62) (or equivalently $Pbnm$ in the $\mathbf{cab}$ setting~\cite{ITA_2016}).
The structure consists of bilayers of edge-sharing MoO$_6$ octahedra, and the bilayers attract each other by the vdW interaction.
There are three symmetrically inequivalent O sites in $\alpha$-MoO\textsubscript{3}.
The oxygen at the terminal site (O\textsubscript{t}) is bound to only one Mo atom and faces the vdW gap.
The oxygen at the assymetric site (O\textsubscript{a}) is bound to two Mo atoms with different Mo--O bond lengths.
The oxygen at the symmetric site (O\textsubscript{s}) is bound to three Mo atoms, among which two Mo--O bonds lie out of the reflection plane and are symmetrically equivalent.
Mo as well as all the symmetrically inequivalent O atoms belong to distinct $4c$ Wyckoff positions.

\subsection{H adsorption locations\label{sec:H_sites}}

\begin{figure}
    \centering
    \includegraphics[width=\linewidth]{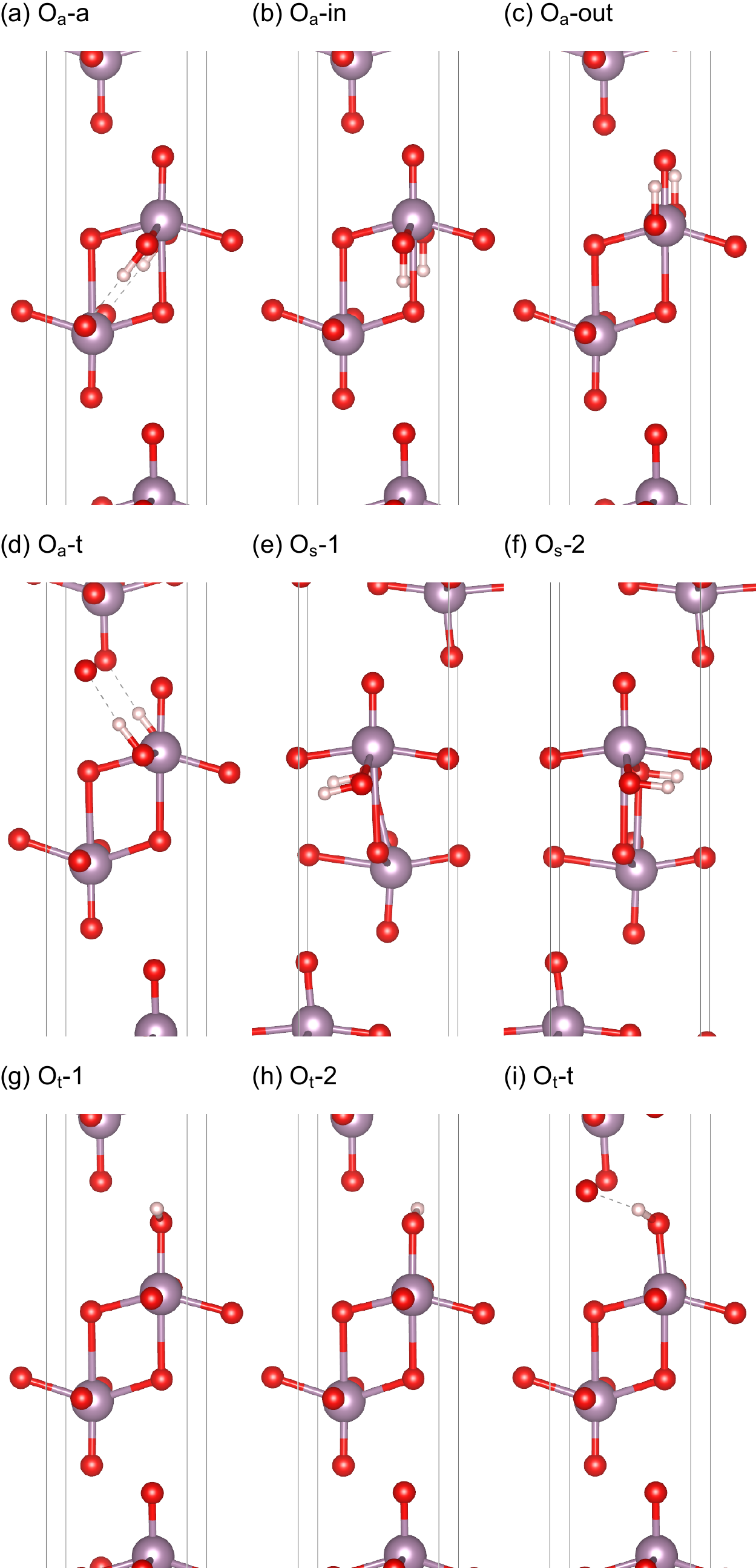}
    \caption{H adsorption sites in $\alpha$-MoO\textsubscript{3} investigated in the present study. Here we show the optimized structures in bulk H$_{0.25}$MoO\textsubscript{3} using the SCAN functional.}
    \label{fig:adsorption_sites}
\end{figure}

\begin{figure}
    \centering
    \includegraphics[width=\linewidth]{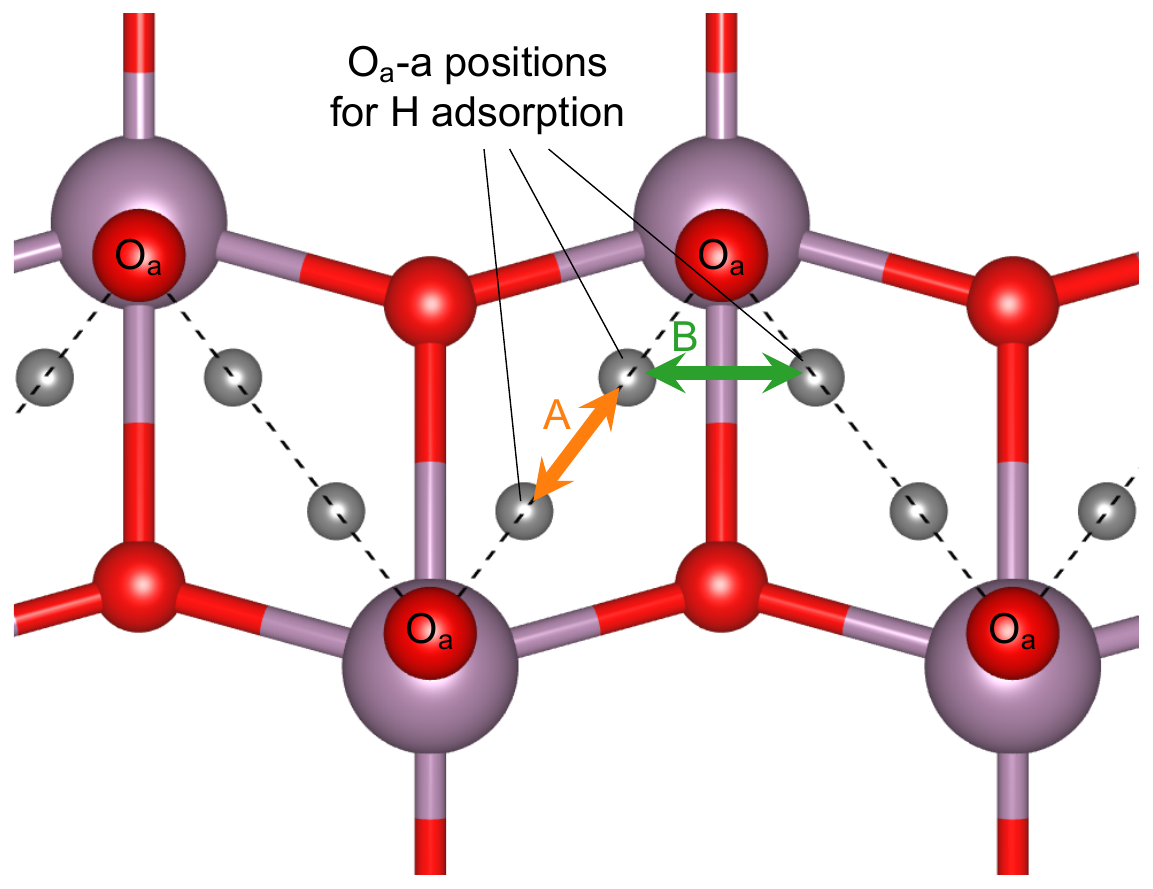}
    \caption{Zig-zag chain of O\textsubscript{a} atoms in the intrabilayer region of $\alpha$-MoO\textsubscript{3}.
    Gray spheres represent the O\textsubscript{a}-a positions potentially occupied by H atoms, and paths A and B are diffusion paths of H atoms between the O\textsubscript{a}-a positions investigated in the present study (cf.~\cref{sec:activation_energy_bulk}).}
    \label{fig:zig-zag}
\end{figure}

An H atom may be adsorbed onto one of the three symmetrically inequivalent O sites in $\alpha$-MoO$_{3}$.
Further, even for the same O site, H can sit at various locations, as exemplified in \cref{fig:adsorption_sites}.
Specifically, for the O\textsubscript{a} site, the H atom can be located on the O\textsubscript{a}--O\textsubscript{a} zig-zag chain in the intrabilayer region (see \cref{fig:zig-zag}), a location we refer to as O\textsubscript{a}-a.
It is also possible that the H atom at the O\textsubscript{a} site resides in the vdW gap and is directed to the O\textsubscript{t} site of the next bilayer, a location we refer to as O\textsubscript{a}-t.
We can further consider the ideal H adsorption onto the O\textsubscript{a} site within the reflection plane. In this case, H can be located inside and outside of the intrabilayer region, which we refer to as O\textsubscript{a}-in and O\textsubscript{a}-out, respectively.

% Note that, during the structural optimization, the O\textsubscript{a}-in and the O\textsubscript{a}-out sites were symmetrically restricted on the reflection plane. Likely these positions are the saddle points on the potential energy hypersurface rather than the dynamically stable sites.

%%%%%
\subsection{Review on previous \textit{ab initio} studies\label{sec:review}}
%%%%%

\begin{table*}[htb]
\centering
\caption{\textit{Ab initio} H adsorption energies (in eV/(H atom)) on the MoO$_3$ surface as reported in previous studies.
For the H adsorption at the O\textsubscript{a} site, we have identified the H position from the visualizations provided in the articles.
Note that the values referenced with respect to H\textsubscript{2} (\cref{eq:adsorption_energy_H2}) \cite{Mei_JPCC_2011_Density,Yang_JPCC_2012_First,Shetty_JPCC_2017_Computational} are shifted to be consistent with the H-referenced values (\cref{eq:adsorption_energy_H}) by the experimental H\textsubscript{2} atomization energy without zero-point energy (ZPE), \SI{2.374}{\electronvolt/(H~atom)}~\cite{Irikura_JPCRD_2007_Experimental,*Irikura_JPCRD_2009_Erratum} (cf.~Supporting Information).}
% \footnotetext[1]{Not explicitly stated.}
\footnotetext[1]{\citet{Yang_JPCC_2012_First} \textit{a priori} assumed the O\textsubscript{t} site as energetically the most favorable one for H with referring to \citet{Sha_JPCC_2009_Hydrogen}, and the H adsorption energies for a single H atom at the O\textsubscript{s} and the O\textsubscript{a} sites were not given.}
\label{tab:adsorption_energies}
\robustify\bfseries
\begin{tabular}{ccccH*{4}{S[detect-weight,table-format=4.3]}}
\toprule
&
Year &
Code &
XC &
Coverage (\si{\percent}) &
{O$_\mathrm{t}$} &
{O$_\mathrm{s}$} &
{O$_\mathrm{a}$-out} &
{O$_\mathrm{a}$-a} \\
\midrule
\citet{Chen_JCP_1998_density}                & 1998 & CASTEP & LDA  &  100 & \bfseries -3.39 & -2.77 & -3.13 & {N/A} \\
\citet{Chen_JPCC_2008_Mechanisms}            & 2008 & VASP   & PW91 &   25 & -2.45 & -2.10 & -2.67 & \bfseries -2.91 \\
\citet{Mei_JPCC_2011_Density}                & 2011 & VASP   & PBE  &   25 & \bfseries -3.60 & -2.70 & -2.24 & {N/A} \\
\citet{Yang_JPCC_2012_First}\footnotemark[1] & 2012 & VASP   & PBE  &  100 & \bfseries -2.78 & {N/A} & {N/A} & {N/A} \\
\citet{Shetty_JPCC_2017_Computational}       & 2017 & CP2K   & PBE  & 6.25 & -2.70 & -2.68 & \bfseries -3.24 & {N/A} \\
\bottomrule
\end{tabular}
% \citet{Chen_JACS_2001_Chemical}        & ACRES  & LDA  & H\footnotemark[1] &   25 & -4.4  & -2.0  & -4.1  & {N/A} \\  % Their also have an O vacancy together and hence not fair comparison.
%\citet{Chen_JPCC_2008_Mechanisms}      & VASP   & PW91 & H                 &   25 & -2.45 & -2.10 & -2.67 & -2.91 \\
% \citet{Mei_JPCC_2011_Density}          & VASP   & PBE  & H$_2$             &   25 & -1.23 & -0.33 & +0.13 & {N/A} \\
% \citet{Yang_JPCC_2012_First}           & VASP   & PBE  & H$_2$             &  100 & -0.41 & {N/A} & {N/A} & {N/A} \\
% \citet{Shetty_JPCC_2017_Computational} & CP2K   & PBE  & H$_2$             & 6.25 & -0.33 & -0.31 & -0.87 & {N/A} \\
\end{table*}

While there are already several \textit{ab initio} reports for the adsorption of an H atom on $\alpha$-MoO$_3$, the conclusions are often inconsistent among each other, both qualitatively and quantitatively.
The most significant inconsistency regards the O site claimed to be the energetically most favorable for H to be adsorbed.
Some studies~\cite{Chen_JPCC_2008_Mechanisms,Shetty_JPCC_2017_Computational,Haque_JMCA_2019_Ordered} reported that the O\textsubscript{a} site is energetically more favorable than the other O sites,
consistent with powder neutron diffraction~\cite{Schroeder_ZAAC_1977_Beitraege,Dickens_JSSC_1979_Elastic}
% Schroeder_ZAAC_1977_Beitraege: H0.5MoO3
% Dickens_JSSC_1979_Elastic: D0.36MoO3
and nuclear magnetic resonance~\cite{Ritter_BBPC_1982_Quasi,Ritter_JCP_1985_Structure}
% Ritter_BBPC_1982_Quasi: H0.35MoO3
% Ritter_JCP_1985_Structure: H0.2MoO3, H0.35MoO3
for bulk H$_x$MoO$_{3}$ with $x \le 0.5$.
Other \textit{ab initio} reports~\cite{Chen_JCP_1998_density,Chen_JACS_2001_Chemical,Sha_JPCC_2009_Hydrogen,Mei_JPCC_2011_Density}, in contrast, claimed that the O\textsubscript{t} site is more favorable than the O\textsubscript{a} site.

\cref{tab:adsorption_energies} summarizes the \textit{ab initio} H adsorption energies on the MoO\textsubscript{3} surface reported previously.
As mentioned in \cref{sec:motivation}, from experiments H is expected to reside in the intrabilayer region and to be bonded to the O\textsubscript{a} site for a relatively low H content of $x \le 0.5$ in bulk H$_x$MoO$_3$.
This H location has, however, been rarely investigated in previous \textit{ab initio} simulations, except for~\citet{Chen_JPCC_2008_Mechanisms}, which is likely one of the major reasons for the inconsistency among previous \textit{ab initio} studies as well as with experiments.
Another possible reason is the difference in the utilized exchange--correlation functional, i.e., specifically, the difference between the local density approximation (LDA) and the generalized gradient approximation (GGA).
This possibility was actually pointed out by \citet{Chen_JPCC_2008_Mechanisms}, but no proof was given at that time.

The significant inconsistencies among previous \textit{ab initio} studies prevent us from referring to the reported values with confidence.
In the present study, we explain the inconsistencies with systematic calculations and provide reliable values of the H adsorption energies on MoO\textsubscript{3}.

%%%%%
\section{Computational Details\label{sec:computational_details}}
%%%%%

\subsection{H adsorption energy}

The H adsorption energy may be computed with reference to either an H atom or an H$_2$ molecule.
When we take an H atom as a reference, the adsorption energy for MoO$_3$ is computed as
\begin{align}
    \Delta E =
    \frac{1}{x}\left[E(\mathrm{H_\mathit{x}MoO_3}) - \left(E(\mathrm{MoO_3}) + x E(\mathrm{H})\right)\right].
    \label{eq:adsorption_energy_H}
\end{align}
When we take an H$_2$ molecule as a reference, the adsorption energy is computed as~\cite{Note1}
% Sha_JPCC_2009_Hydrogen
\begin{align}
    \Delta E =
    \frac{1}{x}\left[E(\mathrm{H_\mathit{x}MoO_3}) - \left(E(\mathrm{MoO_3}) + \frac{x}{2}E(\mathrm{H_2})\right)\right].
    \label{eq:adsorption_energy_H2}
\end{align}
In the present study, we computed the H adsorption energy with reference to an H atom, i.e., based on \cref{eq:adsorption_energy_H}.

%%%%%%%%%%%%%%%%%%%%%%%%%%%%%%%%%%%%%%%%
\subsection{Electronic-structure calculations\label{sec:electronic_structures}}
%%%%%%%%%%%%%%%%%%%%%%%%%%%%%%%%%%%%%%%%

The plane-wave basis projector augmented wave (PAW) method
\cite{Bloechl_PRB_1994_Projector}
% P. E. Bl\”{o}chl, Phys. Rev. B 50, 17953 (1994)
was employed in the framework of density functional theory (DFT)
as implemented in the VASP code
\cite{Kresse_JNS_1995_Ab,Kresse_CMS_1996_Efficiency,Kresse_PRB_1999_ultrasoft}.
% G. Kresse, J. non-Cryst. Solids 192-193, 222 (1995),
% G. Kresse and J. Furthmüller, Comput. Mater. Sci. 6, 15 (1996),
% G. Kresse and D. Joubert, Phys. Rev. B 59, 1758 (1999),
The plane-wave cutoff energy was set to \SI{500}{\electronvolt},
and the Methfessel--Paxton scheme
\cite{Methfessel_PRB_1989_High}
with a smearing width of \SI{0.1}{\electronvolt} was employed for the sampling of the Brillouin zone.
The 4s4p4d5s, 2s2p, and 1s orbitals of Mo, O, and H, respectively, were treated as the valence states.
% while the other electrons were kept frozen.  % PAW
Total energies were minimized until they were converged to within \SI{1e-5}{\electronvolt} per simulation cell for each ionic step.

We considered several exchange--correlation functionals to investigate their impact on the results.
LDA
\cite{Ceperley_PRL_1980_Ground,Perdew_PRB_1981_Self}  % in the Perdew--Zunger form
and GGA in the Perdew--Burke--Ernzerhof (PBE) form
\cite{Perdew_PRL_1996_Generalized}  % J. P. Perdew, K. Burke, and M. Ernzerhof, Phys. Rev. Lett. 77, 3865 (1996)
were used in previous \textit{ab initio} studies and therefore also considered in the present study. These functionals, however, do not consider the vdW interaction.
While there are various approaches to include the vdW interaction in a DFT exchange--correlation functional, \citet{Peelaers_JPCM_2014_First} reported
that the semiempirical DFT-D2 functional proposed by \citet{Grimme_JCC_2006_Semiempirical} in combination with PBE shows reasonable agreements for lattice parameters with experimental values, at reasonable computational costs.
We therefore considered the PBE-D2 functional.
Furthermore, the strongly constrained and appropriately normed (SCAN) meta-GGA~\cite{Sun_PRL_2015_Strongly,Sun_NC_2016_Accurate} was considered, which accurately predicts electronic, thermodynamic, and electronic properties of various systems, even those with vdW interactions.

% \textcolor{C0}{[2021-07-28: For vdW PW91 might behave similarly to LDA: See~\cite{Gerber_JCP_2007_London,Murray_JCTC_2009_Investigation} and some References in \cite{Gerber_JCP_2007_London} for PW91 error.]}

To determine the lattice parameters, bulk MoO$_3$ was computed employing a 16-atom unit cell.
The Brillouin zone was sampled by a $\Gamma$-centered \num{2x6x6} $k$-point mesh.
Cell volume, cell shape, and internal atomic positions were optimized so that the forces on atoms and the stress components on the unit cell became less than \SI{5e-3}{\electronvolt/\angstrom} and \SI{2.5e-4}{\electronvolt/\angstrom{}^3}, respectively.
The structure optimization was repeated several times to be consistent with the given energy cutoff.

To model the MoO$_3$ surface, slab models with \numlist{1x1;2x2;3x3} expansions for the in-plain directions and with a vacuum region in the out-of-plain direction were used.
Three different numbers of MoO$_3$ bilayers (\numlist{1;2;3}) and three vacuum region thicknesses (\SIlist{12;14;16}{\angstrom}) were considered.
The Brillouin zones were sampled by $k$-point meshes consistent with the bulk calculations for the in-plane directions, and with one $k$-point for the out-of-plane direction.
Internal atomic positions of the MoO$_3$ were optimized first without H with the same force convergence criterion as the bulk calculations and with keeping the cell volume and the cell shape consistent with the optimized bulk value.

For each optimized bulk and surface model of MoO$_3$, we put an H atom and reoptimized the internal atomic coordinates with fixed cell volume and cell shape.
The force convergence criterion was set to \SI{2e-2}{\electronvolt/\angstrom}. 
All the calculations including an H atom were performed considering spin polarization.
In the bulk, we considered the nine H positions shown in \cref{fig:adsorption_sites}, while on the surface we considered seven H positions excluding O\textsubscript{a}-t and O\textsubscript{t}-a, which involve two bilayers.
Transition states of the H diffusion were optimized using the improved dimer method~\cite{Heyden_JCP_2005_Efficient}.

\section{Results and Discussion\label{sec:results_and_discussion}}

\subsection{Structural parameters of bulk MoO\texorpdfstring{\textsubscript{3}}{3}\label{sec:bulk}}

\begin{table*}[tb]
    \centering
    \caption{Lattice parameters and Mo--O bond lengths (in \si{\angstrom}) of the bulk $\alpha$-MoO$_3$ obtained in the present \textit{ab initio} simulations.
    Experimental values of \citet{Kihlborg_AK_1963_Least} are also shown for comparison.}
    \label{tab:lattice_parameters}
    \begin{tabular}{
    c
    S[table-format=2.4,table-column-width=1.25cm]
    S[table-format=1.4,table-column-width=1.25cm]
    S[table-format=1.4,table-column-width=1.25cm]
    *5{S[table-auto-round,table-format=1.3,table-column-width=1.125cm]}
}
\toprule
                                            &  {$a$} &  {$b$} &  {$c$} &  {Mo--O\textsubscript{t}} &  \multicolumn{2}{c}{Mo--O\textsubscript{a}} &  \multicolumn{2}{c}{Mo--O\textsubscript{s}} \\
\midrule
                                        LDA & 13.575 &  3.678 &  3.816 &                    1.674835 &                    1.765862 &                    2.079242 &                    1.926466 &                    2.341379 \\
                                        PBE & 15.988 &  3.691 &  3.936 &                    1.680488 &                    1.759894 &                    2.199591 &                    1.947490 &                    2.470634 \\
                                     PBE-D2 & 13.932 &  3.693 &  3.909 &                    1.681602 &                    1.756213 &                    2.181259 &                    1.942943 &                    2.412957 \\
\textbf{SCAN} & 14.044 &  3.681 &  3.907 &                    1.673411 &                    1.745301 &                    2.185417 &                    1.939807 &                    2.379557 \\
\midrule Exp.~\cite{Kihlborg_AK_1963_Least} & 13.855 & 3.6964 & 3.9628 &                    1.671000 &                    1.734000 &                    2.251000 &                    1.948000 &                    2.332000 \\
\bottomrule
\end{tabular} \end{table*}

\cref{tab:lattice_parameters} shows the \textit{ab initio} lattice parameters of bulk $\alpha$-MoO\textsubscript{3} obtained in the present study.

For the in-plane lattice parameters $b$ and $c$, all the exchange--correlation functionals except for LDA show a reasonable agreement with experiment with an error of only up to \SI{1.5}{\percent}. LDA underestimates $b$ and $c$ strongly, with an error up to \SI{3.7}{\percent}, demonstrating its overbinding nature~\cite{Haas_PRB_2009_Calculation,*Haas_PRB_2009_Erratum,Grabowski_PRB_2007_Ab,Hickel_JPCM_2011_Advancing}.

For the out-of-plane lattice parameter $a$, PBE largely overestimates experiment by \SI{15}{\percent}, as also reported in previous studies~\cite{Peelaers_JPCM_2014_First,Zhang_JCP_2019_Surface}, clearly indicating the failure of PBE in describing the vdW interaction.
LDA seems to show a good agreement with experiments, but as discussed frequently previously~\cite{Harris_PRB_1985_Simplified,Bjoerkman_JPCM_2012_Are}, this is just due to error cancellation because LDA assigns an unphysically long range to exchange interactions and not because LDA captures the vdW interaction.
In contrast, PBE-D2 and SCAN, both of which take the vdW interaction into account, predict $a$ with an error of only up to \SI{1.4}{\percent}, consistent with previous studies~\cite{Peelaers_JPCM_2014_First,Zhang_JCP_2019_Surface}.

For the Mo--O bond lengths, in most cases all the investigated exchange--correlation functionals show good agreement with experimental values, but there are a few exceptions.
Specifically, LDA shows an error of \SI{-7.6}{\percent} for the longer Mo--O\textsubscript{a} bond, and PBE and PBE-D2 show errors of \SIlist{+5.9;+3.5}{\percent}, respectively, for the longer Mo--O\textsubscript{s} bond.
The SCAN functional does not show errors larger than \SI{3}{\percent} for any of the Mo--O bond lengths and thus best predicts the experimental structure, which is consistent with the results of \citet{Zhang_JCP_2019_Surface}.

%%%%%
\subsection{Surface energy of MoO\texorpdfstring{\textsubscript{3}}{3}}
%%%%%

\begin{figure}[tp]
    \centering
    \includegraphics[width=\linewidth]{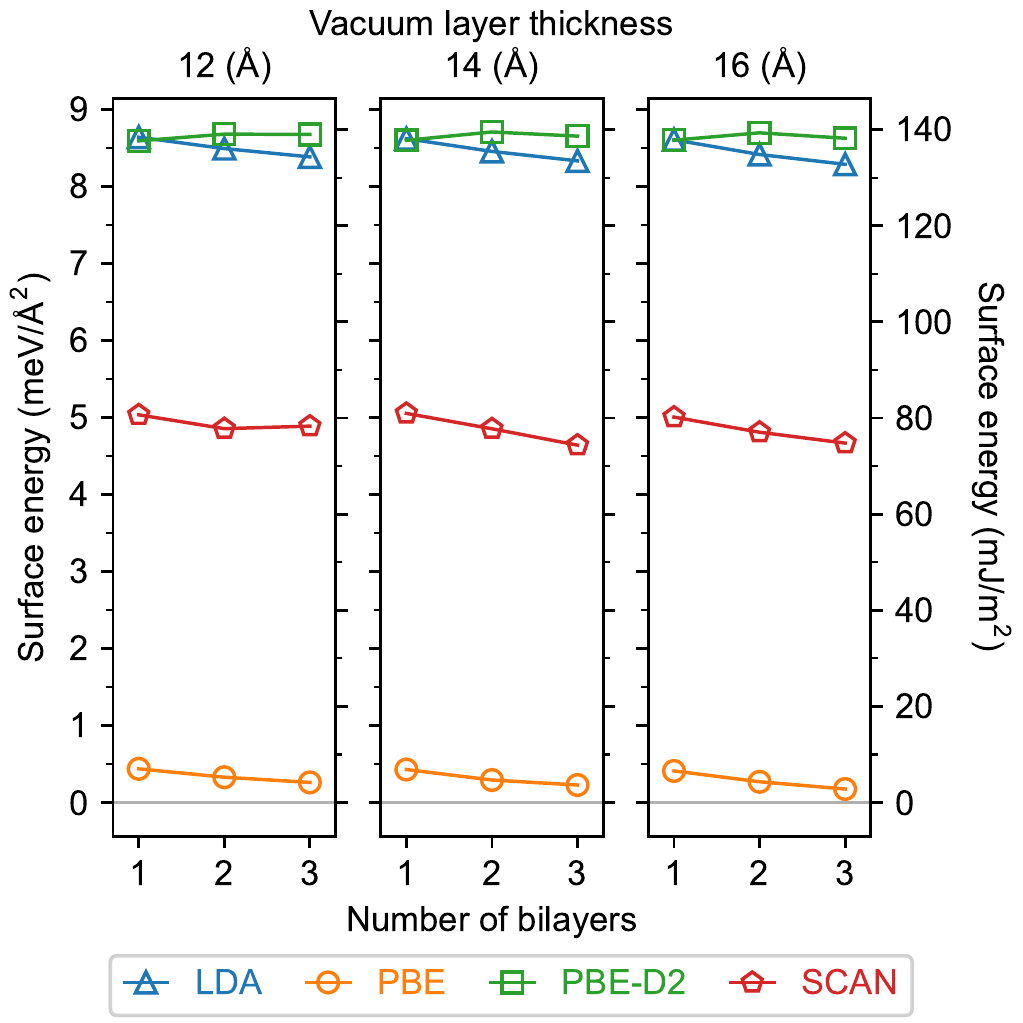}
    \caption{Surface energies of $\alpha$-MoO\textsubscript{3} obtained in the present \textit{ab initio} calculations.}
    \label{fig:surface_energies}
\end{figure}

\cref{fig:surface_energies} shows the \textit{ab initio} surface energies of $\alpha$-MoO$_3$ obtained in the present study.

The values are almost independent of the vacuum region thickness, indicating that even the smallest thickness (\SI{12}{\angstrom}) would be enough to avoid bilayer interactions between mirror images of the slab models.
We hereafter focus on the results obtained using a vacuum layer thickness of \SI{14}{\angstrom}. The obtained surface energies are likewise almost independent of the number of bilayers, implying that the surface--surface interaction is marginal even for one bilayer.

PBE-D2 shows a surface energy of about \SI{8.7}{\milli\electronvolt/\angstrom{}^2}, while PBE shows a much smaller value of about \SI{0.2}{\milli\electronvolt/\angstrom{}^2}, indicating a very weak interaction between the bilayers due to the absence of the vdW interaction in PBE. SCAN shows a surface energy of about \SI{4.6}{\milli\electronvolt/\angstrom{}^2}, in between the values of PBE and PBE-D2.
These values are in good agreement with a previous \textit{ab initio} study~\cite{Zhang_JCP_2019_Surface}.
%%%%%
LDA seems to show similar values to those of PBE-D2. As discussed in \cref{sec:bulk}, however, by its nature LDA cannot correctly simulate the vdW interaction~\cite{Harris_PRB_1985_Simplified}, and therefore the agreement should be attributed to a coincidence~\cite{Bjoerkman_JPCM_2012_Are}.

%%%%%
\subsection{H adsorption energy in bulk MoO\texorpdfstring{\textsubscript{3}}{3}\label{sec:absorption_energy_bulk}}
%%%%%

\begin{figure*}
    \centering
    \includegraphics[width=\linewidth]{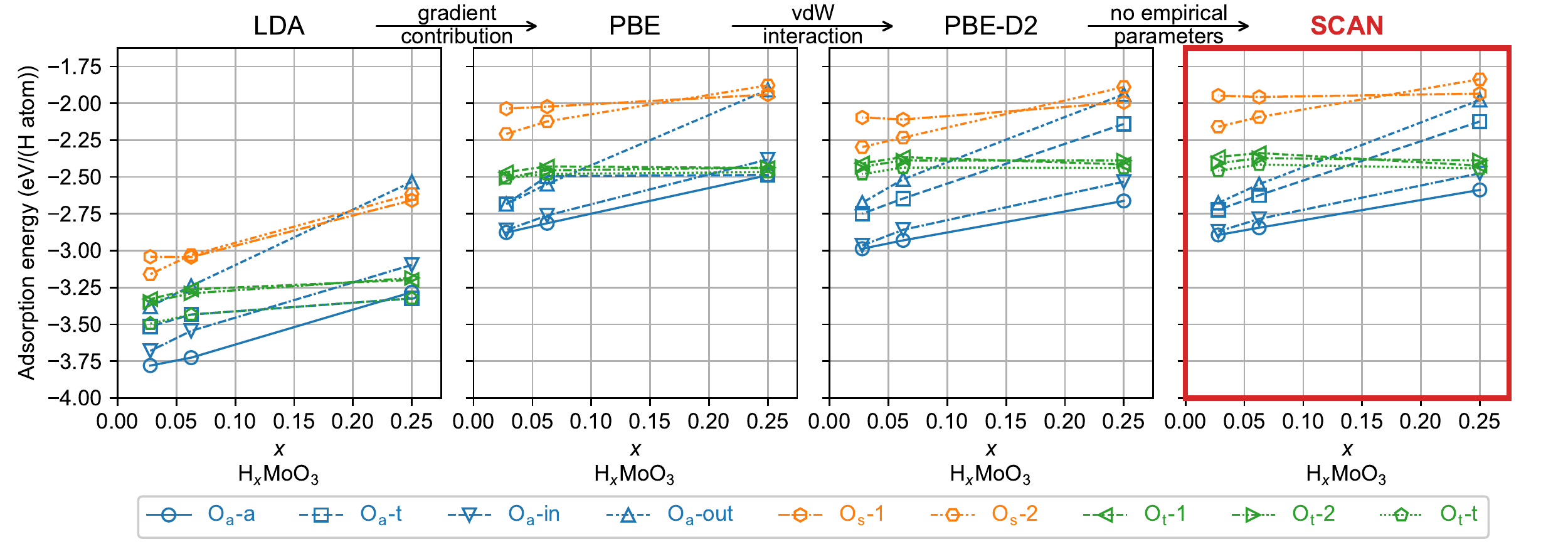}
    \caption{H adsorption energies in bulk MoO\textsubscript{3} with H as a reference (\cref{eq:adsorption_energy_H}).
    The lines are guides for the eyes.}
    \label{fig:absorption_energy_bulk}
\end{figure*}

\begin{figure*}
    \centering
    \includegraphics[width=\linewidth]{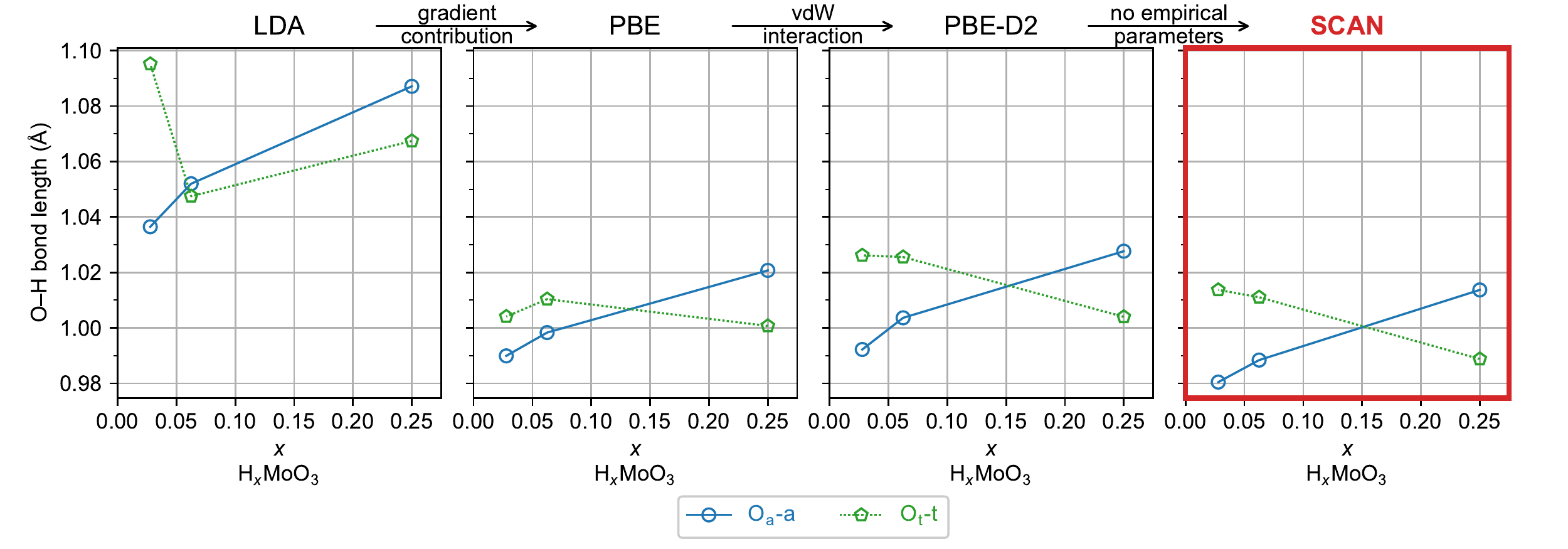}
    \caption{O--H bond lengths in bulk MoO\textsubscript{3} for the O\textsubscript{a}-a and the O\textsubscript{t}-t positions.
    The lines are guides for the eyes.}
    \label{fig:O-H_bulk}
\end{figure*}

\cref{fig:absorption_energy_bulk} shows the \textit{ab initio} computed H adsorption energies in bulk $\alpha$-MoO\textsubscript{3} for various computational conditions.

Among the investigated exchange--correlation functionals, LDA shows much lower adsorption energies than the other exchange--correlation functionals.
In general, for various properties, PBE and SCAN show better agreements with experiments than LDA~\cite{Sun_PRL_2015_Strongly,Sun_NC_2016_Accurate}.
The much lower H adsorption energies in LDA demonstrate the known overbinding nature of the LDA~\cite{Haas_PRB_2009_Calculation,*Haas_PRB_2009_Erratum,Grabowski_PRB_2007_Ab,Hickel_JPCM_2011_Advancing}.
It should be however noted that the \emph{relative} H adsorption energies among the investigated H positions predicted in LDA are still mostly similar to the other exchange--correlation functionals.
PBE-D2 and SCAN, both of which are expected to be able to simulate the vdW interaction, show very similar trends, even quantitatively.

Except for LDA at $x = 0.25$, the O\textsubscript{a}-a position is energetically the most favorable site for H regardless of the exchange--correlation functional and the H content.
This is consistent with experiments for $x < 0.25$~\cite{Schroeder_ZAAC_1977_Beitraege,Dickens_JSSC_1979_Elastic,Ritter_BBPC_1982_Quasi,Ritter_JCP_1985_Structure}, where H resides along the zig-zag chains of the O\textsubscript{a} sites in the intrabilayer region.

At $x = 0.25$, LDA shows that the O\textsubscript{t} site (O\textsubscript{t}-t) is energetically slightly more stable than the O\textsubscript{a} site (O\textsubscript{a}-a). Specifically, for LDA, H adsorption energies are
\SI[round-mode=figures,round-precision=3]{-3.323944430}{\electronvolt} and 
\SI[round-mode=figures,round-precision=3]{-3.282031010}{\electronvolt} at the O\textsubscript{t} and the O\textsubscript{a} sites, respectively, and therefore the O\textsubscript{t} site is \SI{0.04}{\electronvolt} more stable than the O\textsubscript{a} site.
For PBE, H adsorption energies are  
\SI[round-mode=figures,round-precision=3]{-2.466658020}{\electronvolt} and 
\SI[round-mode=figures,round-precision=3]{-2.488096520}{\electronvolt} at the O\textsubscript{t} and the O\textsubscript{a} sites, respectively, and therefore the O\textsubscript{a} site is only \SI{0.02}{\electronvolt} more stable than the O\textsubscript{t} site.
% Furthermore, also in PBE at $x = 0.25$, the O\textsubscript{t}-t and the O\textsubscript{a}-a positions show very similar H adsorption energies with a difference of only \SI{0.02}{\electronvolt}.
Although this small energy difference between the O\textsubscript{t} and the O\textsubscript{a} sites is in line with previous \textit{ab initio} results by \citet{Sha_JPCC_2009_Hydrogen},
% , where H was shown to be \SI{0.03}{\electronvolt}/(H atom) more stable at the O\textsubscript{t} site than at the O\textsubscript{a} site for $x = 0.25$ and with PBE,
it is contradictory to powder neutron diffraction~\cite{Schroeder_ZAAC_1977_Beitraege,Dickens_JSSC_1979_Elastic}
% Schroeder_ZAAC_1977_Beitraege: H0.5MoO3
% Dickens_JSSC_1979_Elastic: D0.36MoO3
and nuclear magnetic resonance~\cite{Ritter_BBPC_1982_Quasi,Ritter_JCP_1985_Structure}
for $x \le 0.5$, which showed that H occupies the O\textsubscript{a} site.
For PBE-D2 and SCAN, in contrast to LDA and PBE, the O\textsubscript{a} site is correctly predicted to be substantially more stable for H than the O\textsubscript{t} site.
Specifically, for PBE-D2, H adsorption energies are
\SI[round-mode=figures,round-precision=3]{-2.438228780}{\electronvolt} and  \SI[round-mode=figures,round-precision=3]{-2.662329710}{\electronvolt} at the O\textsubscript{t} and the O\textsubscript{a} sites, respectively, and therefore the 
O\textsubscript{a} site is \SI{0.22}{\electronvolt} more stable than the O\textsubscript{t} site.
For SCAN, H adsorption energies are 
\SI[round-mode=figures,round-precision=3]{-2.441885550}{\electronvolt} and \SI[round-mode=figures,round-precision=3]{-2.587953230}{\electronvolt} at the O\textsubscript{t} and the O\textsubscript{a} sites, respectively, and therefore the O\textsubscript{a} site is \SI{0.15}{\electronvolt} more stable than the O\textsubscript{t} site.
The poor agreement of LDA and PBE with the experimental observation can be probably ascribed to the lack of the consideration of the vdW interaction in these functionals.
Particularly for PBE, since the interbilayer distance is largely overestimated (cf.~\cref{tab:lattice_parameters}), it may erroneously stabilize the O\textsubscript{t} site, which faces the vdW region, as compared with the O\textsubscript{a} site.
These results clearly demonstrate the importance of exchange--correlation functionals which account for the vdW interaction to correctly predict the most stable H adsorption site.

For the dilute H concentration of $x \le 0.0625$, where the H--H interaction is expected to be negligible, H substantially favors the O\textsubscript{a} site rather than the O\textsubscript{t} site, irrespective of the exchange--correlation functionals.
When the H content increases, the H adsorption energy at the O\textsubscript{a} site becomes less negative while the H adsorption energy at the O\textsubscript{t} site is less sensitive against the H content.
This implies that the O\textsubscript{t} site would be energetically more stable than the O\textsubscript{a} site at further higher H concentrations, which is consistent with experimental observations for hydrogen molybdenum bronze.
Specifically, at low H concentrations ($x \le 0.5$) H is bound to the O\textsubscript{a} site and located on the zig-zag chain in the intrabilayer region~\cite{Schroeder_ZAAC_1977_Beitraege,Dickens_JSSC_1979_Elastic,Ritter_BBPC_1982_Quasi,Ritter_JCP_1985_Structure}, while for higher H contents of $1.55 < x \le 2$, powder neutron diffraction~\cite{Dickens_SSI_1988_crystal,Anne_JPF_1988_Structure}
% Dickens_SSI_1988_crystal: D1.68MoO3
% Anne_JPF_1988_Structure: D1.65MoO3
showed that H is bound to the O\textsubscript{t} sites and resides in the interbilayer region.

% We must note that in Ref.~\cite{Sha_JPCC_2009_Hydrogen} the sign convention of the H adsorption energy is inconsistent between their definition and their values and also their values are also inconsistent between the Tables and their main text, which hinders the survey both qualitatively and quantitatively.
% It should be also noted that \citet{Sha_JPCC_2009_Hydrogen} do not consider the in-plane expansion, and therefore H--H repulsion between mirror images due to the periodic boundary conditions may be large.

Irrespective of the exchange--correlation functional as well as of the H content, the O\textsubscript{a}-a position is the most favorable among the O\textsubscript{a} sites, and the O\textsubscript{t}-t position is the most favorable among the O\textsubscript{t} sites.
\cref{fig:O-H_bulk} shows the O--H bond lengths for these two selected H positions.
LDA shows substantially longer (\SI{>1.03}{\angstrom}) O--H bond lengths compared with the other exchange--correlation functionals.
In SCAN, when the H content increases, the O--H bond lengths become longer and shorter for the O\textsubscript{a}-a and the O\textsubscript{t}-t positions, respectively.
This trend is correlated with the energetic stability of an H atom in \cref{fig:absorption_energy_bulk}.
That is, at the O\textsubscript{a}-a position, H adsorption becomes less favorable with the increase of the H content, together with the elongation of the O--H bond lengths, and the opposite trend is found for the O\textsubscript{t}-t position.

\subsection{H activation energy in bulk MoO\texorpdfstring{\textsubscript{3}}{3}\label{sec:activation_energy_bulk}}

\begin{table}
    \centering
    \caption{H activation energies $\Delta^\ddagger E$ and O--H bond lengths for the transition states of the H diffusion paths A and B shown in \cref{fig:zig-zag} for bulk H$_{0.25}$MoO$_3$.}
    \label{tab:activation_energies}
\begin{tabular}{*5{P{0.18\linewidth}}}
\toprule
{} & \multicolumn{2}{c}{$\Delta^\ddagger E$ (\si{\electronvolt/(H~atom))}} & \multicolumn{2}{c}{O--H (\si{\angstrom})} \\
\cmidrule(lr){2-3}\cmidrule(lr){4-5}
{} &                                                 A &      B &                     A &      B \\
\midrule
LDA    &                                             0.011 &  0.185 &                 1.231 &  1.007 \\
PBE    &                                             0.116 &  0.107 &                 1.248 &  0.993 \\
PBE-D2 &                                             0.091 &  0.131 &                 1.240 &  0.994 \\
\bf SCAN &                                             0.146 &  0.113 &                 1.236 &  0.984 \\
\bottomrule
\end{tabular}
\end{table}

Having verified that the O\textsubscript{a}-a position is the most favorable for H in bulk $\alpha$-MoO\textsubscript{3}, we now focus on the H activation energies in bulk $\alpha$-MoO\textsubscript{3} for H diffusion between the O\textsubscript{a}-a positions. Specifically, we consider the intralayer H diffusion paths A and B shown in \cref{fig:zig-zag}.
\cref{tab:activation_energies} shows the \textit{ab initio} computed H activation energies.
Except for path A computed with LDA, the obtained H activation energies are \SIrange{0.1}{0.2}{\electronvolt/(H~atom)}~\cite{Note2}.
This is in good agreement with previous experimental values; the value by \citet{Slade_JSSC_1980_NMR} (\SI{0.11}{\electronvolt/(H~atom)} for $x = 0.36$) is almost in perfect agreement with our \textit{ab initio} values, and the values of \citet{Ritter_BBPC_1982_Quasi} (\SI{0.31}{{\electronvolt/(H~atom)}} for $x = 0.35$) and \citet{Ritter_JCP_1985_Structure} (\SIrange{0.28}{0.31}{\electronvolt/(H~atom)} for $x$ in 0.2--0.35) are just slightly higher than our \textit{ab initio} values.

While the O--H bond lengths are similar among the exchange--correlation functionals for both paths A and B, the trends of the H activation energies are different.
In PBE and SCAN, path A shows lower H activation energies, while LDA and PBE-D2 show the opposite trend.
From the known successes of the SCAN functional~\cite{Sun_PRL_2015_Strongly,Sun_NC_2016_Accurate} as well as the best geometrical agreement with experiments for bulk MoO$_3$ in the present study (cf.~\cref{tab:lattice_parameters}), we expect that SCAN provides more reliable results.

%%%%%
\subsection{H adsorption energy on the MoO\texorpdfstring{\textsubscript{3}}{3} surface}
%%%%%

\begin{figure*}
    \centering
    \includegraphics[width=\linewidth]{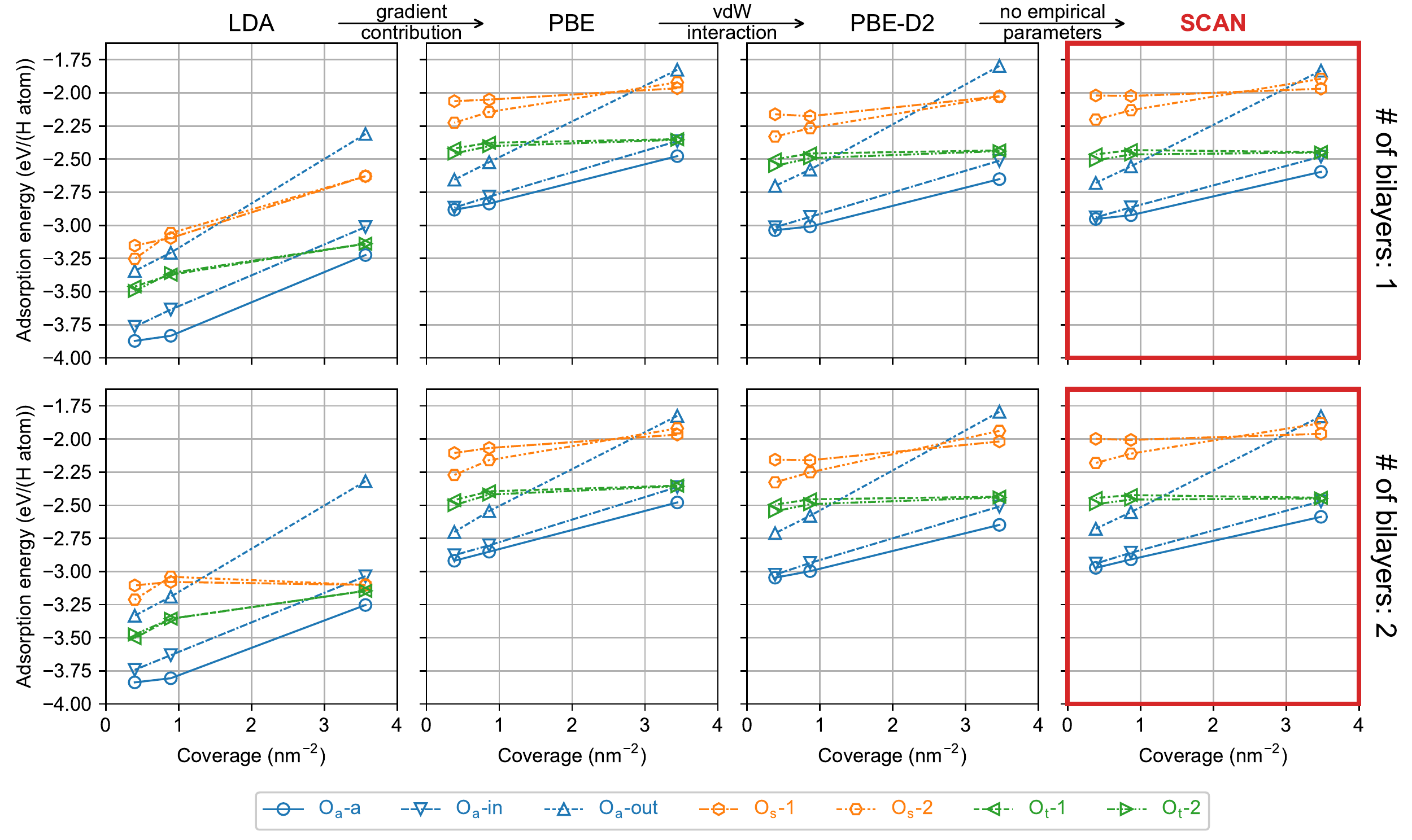}
    \caption{H adsorption energies on the MoO\textsubscript{3} surface with H as a reference (\cref{eq:adsorption_energy_H}). The lines are guides for the eyes.}
    \label{fig:adsoprtion_energies}
\end{figure*}

\cref{fig:adsoprtion_energies} shows the \textit{ab initio} computed H adsorption energies on the MoO\textsubscript{3} surface for various computational conditions.
The values are almost independent of the number of bilayers (top vs.~bottom panels in each column) for all of the investigated exchange--correlation functionals and for all the H coverages, meaning that the impact of H is limited to within one bilayer.

Most of the trends are similar to the H adsorption in bulk MoO\textsubscript{3}.
In particular, LDA gives much more negative H adsorption energies than the other functionals, which can be again ascribed to the overbinding nature of LDA~\cite{Haas_PRB_2009_Calculation,*Haas_PRB_2009_Erratum,Grabowski_PRB_2007_Ab,Hickel_JPCM_2011_Advancing}.
Previous \textit{ab initio} studies using LDA~\cite{Chen_JCP_1998_density}
% Chen_JACS_2001_Chemical  % with O vacancies
likewise showed significantly more negative H adsorption energies than GGA~\cite{Chen_JPCC_2008_Mechanisms,Mei_JPCC_2011_Density,Yang_JPCC_2012_First,Shetty_JPCC_2017_Computational}, even taking the difference of the energy reference (\cref{eq:adsorption_energy_H,eq:adsorption_energy_H2}) into account (cf.~Appendix), as shown in \cref{tab:adsorption_energies}.

The O\textsubscript{a}-a site is energetically the most favorable among the investigated H positions for all the investigated exchange--correlation functionals and for all the investigated H coverages.
As described in \cref{sec:absorption_energy_bulk}, also in bulk MoO\textsubscript{3} the O\textsubscript{a}-a site is the energetically most favorable one when $x \le 0.0625$, which is supported by
powder neutron diffraction~\cite{Schroeder_ZAAC_1977_Beitraege,Dickens_JSSC_1979_Elastic}
% Schroeder_ZAAC_1977_Beitraege: H0.5MoO3
% Dickens_JSSC_1979_Elastic: D0.36MoO3
and by nuclear magnetic resonance~\cite{Ritter_BBPC_1982_Quasi,Ritter_JCP_1985_Structure}
for bulk H$_x$MoO\textsubscript{3} with $x \le 0.5$.

It should be pointed out that, if we consider only the O\textsubscript{a}-out position and ignore the energetically more favorable O\textsubscript{a}-a or O\textsubscript{a}-in positions, as done in Refs.~\cite{Chen_JCP_1998_density,Mei_JPCC_2011_Density,Yang_JPCC_2012_First,Shetty_JPCC_2017_Computational}, the O$_\mathrm{t}$ site (green in \cref{fig:adsoprtion_energies}) can be obtained as the energetically most preferable site in some cases.
Particularly for LDA, regardless of the H coverage, the O\textsubscript{t} site would be obtained as the energetically most stable site, if only the O\textsubscript{a}-out position was considered for the O\textsubscript{a} site, which is indeed the case in an early \textit{ab initio} study using LDA~\cite{Chen_JCP_1998_density,Note3}.
Our present results reveal that the previous erroneous results originate from the fact that 1) LDA was used and 2) the O\textsubscript{a}-a H adsorption site was ignored, which is energetically more favorable than O\textsubscript{a}-out.

Contrary to the present results (\cref{fig:adsoprtion_energies}), a previous \textit{ab initio} study by \citet{Mei_JPCC_2011_Density} reported that even in PBE the O\textsubscript{t} site is energetically the most favorable for H at a coverage of \SI{\approx1}{\per\nano\meter\squared}, although here and in Ref.~\cite{Mei_JPCC_2011_Density} the same \textit{ab initio} code (VASP) was used.
\citet{Mei_JPCC_2011_Density} considered only the O\textsubscript{a}-out position for H and did not consider the O\textsubscript{a}-a and the O\textsubscript{a}-in position~\cite{Note4}.
According to our results, the O\textsubscript{a}-a and the O\textsubscript{a}-in positions are more than \SI{0.2}{\electronvolt} lower in H adsorption energy than the O\textsubscript{a}-out position in PBE and at an H coverage of \SI{\approx1}{\per\nano\meter\squared}, but this does not fully account for their results showing a much more stable O\textsubscript{t} site.
Other detailed computational conditions are also different between \citet{Mei_JPCC_2011_Density} and us.
Specifically, we used a tighter force convergence criterion of \SI{0.02}{\electronvolt/\angstrom} for H-including calculations (\SI{0.05}{\electronvolt/\angstrom} in Ref.~\cite{Mei_JPCC_2011_Density}), a higher energy cutoff of \SI{500}{\electronvolt} (\SI{400}{\electronvolt} in Ref.~\cite{Mei_JPCC_2011_Density}), and a thicker vacuum region of \SI{14}{\angstrom} (\SI{12}{\angstrom} in Ref.~\cite{Mei_JPCC_2011_Density}).
Furthermore, \citet{Mei_JPCC_2011_Density} fixed the in-plane lattice parameters to the experimental values for the bulk rather than relaxing them and also fixed the atoms in the bottom bilayers in their models including two bilayers which were not optimized.
Such detailed differences are, however, not expected to lead to such a contradictory conclusion.
We therefore attribute the contradiction to potential erroneous analyses in the previous research~\cite{Mei_JPCC_2011_Density}.

In \cref{fig:adsoprtion_energies}, the H adsorption energy at the O\textsubscript{a}-a site becomes less negative when the H coverage becomes higher, similarly to the H adsorption in bulk MoO\textsubscript{3} (cf.~\cref{fig:absorption_energy_bulk}), which should be ascribed to the H--H repulsion.
This means that, under the assumption that all the H atoms are adsorbed on the O atoms in the same way, H adsorption becomes energetically less preferable when the amount of H increases.
% This is qualitatively consistent with \citet{Mei_JPCC_2011_Density} and \citet{Yang_JPCC_2012_First}; they used similar computational conditions, except that \citet{Mei_JPCC_2011_Density} and \citet{Yang_JPCC_2012_First} did the \SIlist{25;100}{\percent} H coverages, respectively, and the H adsorption energy reported by \citet{Yang_JPCC_2012_First} is substantially higher (i.e., energetically less stable) than that reported by~\citet{Mei_JPCC_2011_Density}.

%%%%%
\section{Conclusions\label{sec:conclusions}}
%%%%%

\begin{table*}
\caption{H adsorption energies obtained with SCAN as a function of the in-plane supercell expansion. Corresponding H coverages per bilayer (in \si{\per\nano\meter\squared}) are also shown.}
\label{tab:energies_SCAN}
\begin{tabular}{ccc*9{S[table-format=2.3,table-column-width=1.25cm]}}
\toprule
&{Coverage}&{\# of bilayers}&  {O$_\mathrm{a}$-a} &  {O$_\mathrm{a}$-in} &  {O$_\mathrm{a}$-out} &  {O$_\mathrm{a}$-t} &  {O$_\mathrm{s}$-1} &  {O$_\mathrm{s}$-2} &  {O$_\mathrm{t}$-1} &  {O$_\mathrm{t}$-2} &  {O$_\mathrm{t}$-t} \\
\midrule
\num{1x1} &    3.477 &    1 &              -2.596 &               -2.481 &                -1.832 &               {N/A} &              -1.969 &              -1.894 &              -2.447 &              -2.451 &               {N/A} \\
          &          &    2 &              -2.587 &               -2.470 &                -1.828 &               {N/A} &              -1.961 &              -1.881 &              -2.444 &              -2.449 &               {N/A} \\
          &          & bulk &              -2.588 &               -2.475 &                -1.977 &              -2.124 &              -1.934 &              -1.837 &              -2.423 &              -2.388 &              -2.442 \\
\midrule
\num{2x2} &    0.869 &    1 &              -2.923 &               -2.864 &                -2.555 &               {N/A} &              -2.023 &              -2.131 &              -2.432 &              -2.465 &               {N/A} \\
          &          &    2 &              -2.909 &               -2.858 &                -2.552 &               {N/A} &              -2.007 &              -2.110 &              -2.423 &              -2.458 &               {N/A} \\
          &          & bulk &              -2.844 &               -2.785 &                -2.549 &              -2.624 &              -1.957 &              -2.093 &              -2.337 &              -2.372 &              -2.415 \\
\midrule
\num{3x3} &    0.386 &    1 &              -2.952 &               -2.940 &                -2.680 &               {N/A} &              -2.020 &              -2.202 &              -2.464 &              -2.509 &               {N/A} \\
          &          &    2 &              -2.973 &               -2.940 &                -2.677 &               {N/A} &              -1.999 &              -2.181 &              -2.445 &              -2.492 &               {N/A} \\
          &          & bulk &              -2.893 &               -2.870 &                -2.676 &              -2.724 &              -1.948 &              -2.159 &              -2.364 &              -2.406 &              -2.458 \\
\bottomrule
\end{tabular}
 \end{table*}

In the present study, we have performed detailed investigations of the H adsorption on $\alpha$-MoO\textsubscript{3} based on systematic \textit{ab initio} calculations and critical analysis of the existing literature.
Both in the bulk and on the surface of MoO\textsubscript{3}, H prefers to bind to O at the asymmetric site (O\textsubscript{a}) and to reside in the intrabilayer region along the zig-zag chains consisting of the O\textsubscript{a} sites (cf.~\cref{fig:zig-zag}), which is supported by powder neutron diffraction~\cite{Schroeder_ZAAC_1977_Beitraege,Dickens_JSSC_1979_Elastic}
% Schroeder_ZAAC_1977_Beitraege: H0.5MoO3
% Dickens_JSSC_1979_Elastic: D0.36MoO3
and by nuclear magnetic resonance~\cite{Ritter_BBPC_1982_Quasi,Ritter_JCP_1985_Structure}
for bulk H$_x$MoO\textsubscript{3} with $x \le 0.5$.
Previous contradictory \textit{ab initio} results are ascribed to the use of exchange--correlation functionals that either do not take the vdW interaction into account, such as the LDA and the PBE functionals, or to the neglect of H adsorption at the O\textsubscript{a} site \textit{in the intrabilayer region}.
Particularly for bulk MoO\textsubscript{3}, the lack of the vdW interaction in the exchange--correlation functionals is critical, because the distances between the bilayers are overestimated and thus the strength of the H bonding to the O atom at the terminal site (O\textsubscript{t}) is predicted as too strong.
It is clearly advisable to use an exchange--correlation functional that takes the vdW interaction into account, such as the SCAN meta-GGA functional.
Indeed, the SCAN functional offers an H activation energy of \SIrange{0.11}{0.15}{\electronvolt/(H~atom)}, in good agreement with experimental values.
For dilute H content, i.e., $x = 0.0625$ in bulk H$_x$MoO$_3$, the H adsorption energy is predicted as \SI[round-mode=figures,round-precision=3]{-2.893444}{\electronvolt/(H~atom)} at the O\textsubscript{a} site in the intrabilayer region using the SCAN functional.
\cref{tab:energies_SCAN} summarizes the H adsorption energies obtained with SCAN.

\section*{Associated content}

\subsection*{Supporting Information}

The Supporting Information contains the results of H$_2$ atomization energies and zero-point energies.

\begin{acknowledgements}

This project has received funding from the European Research Council (ERC) under the European Union’s Horizon 2020 research and innovation programme (grant agreement No 865855). The authors also acknowledge support by the state of Baden-Württemberg through bwHPC and the German Research Foundation (DFG) through grant no INST 40/467-1 FUGG (JUSTUS cluster). B.G.~acknowledges the support by the Stuttgart Center for Simulation Science (SimTech). D.P.E.~gratefully acknowledges the Deutsche Forschungsgemeinschaft (project number 444948747) for funding.

\end{acknowledgements}

% \bibliographystyle{apsrev4-2}
% \bibliography{bibexport,footnotes}
%apsrev4-2.bst 2019-01-14 (MD) hand-edited version of apsrev4-1.bst
%Control: key (0)
%Control: author (8) initials jnrlst
%Control: editor formatted (1) identically to author
%Control: production of article title (0) allowed
%Control: page (0) single
%Control: year (1) truncated
%Control: production of eprint (1) enabled
%
 
\end{document}

\title{\texorpdfstring{\underline{Supporting Information}}{Supporting Information}\texorpdfstring{\vspace{1em}\\}{ }Comprehensive understanding of H adsorption on MoO\texorpdfstring{\textsubscript{3}}{3} from\texorpdfstring{\\}{ }systematic \textit{ab initio} simulations}

\author{Yuji Ikeda}
\email{yuji.ikeda@imw.uni-stuttgart.de}
\affiliation{Institute for Materials Science, University of Stuttgart, Pfaffenwaldring 55, 70569 Stuttgart, Germany}

\author{Deven Estes}
\affiliation{Institute of Technical Chemistry, University of Stuttgart, Pfaffenwaldring 55, 70569, Stuttgart, Germany}

\author{Blazej Grabowski}
\affiliation{Institute for Materials Science, University of Stuttgart, Pfaffenwaldring 55, 70569 Stuttgart, Germany}

\maketitle

%\tableofcontents

\appendix*

\section{H\texorpdfstring{\textsubscript{2}}{2} atomization energy}

When comparing the H adsorption energies reported in previous \textit{ab initio} studies
% (Table~\ref{tab:adsorption_energies}),
(Table~1 in the main text),
one of the issues is the difference of the definition of the H adsorption energy, i.e., whether we set an H atom
% (Eq.~\eqref{eq:adsorption_energy_H})
(eq~1 in the main text)
or an H\textsubscript{2} molecule
% (Eq.~\eqref{eq:adsorption_energy_H2})
(eq~2 in the main text)
as a reference.
For a fair comparison, we have corrected the values referring to H\textsubscript{2}~\cite{Mei_JPCC_2011_Density,Yang_JPCC_2012_First,Shetty_JPCC_2017_Computational} to refer to H by utilizing the experimental energy difference between H\textsubscript{2} and 2H extracted below.
We have also compared the \textit{ab initio} obtained values with the experimental ones to check the consistency.

% Experimental values are converted in
% /Users/ikeda/Documents/projects/2021/MoO3_DE/references

The \textit{ab initio} calculations of H and H\textsubscript{2} were performed in a \SI{20x20x20}{\angstrom} simulation cell, and only the $\Gamma$ point was used to sample the Brillouin zone.
The atomic distance of H\textsubscript{2} was optimized so that the forces on atoms are less than \SI{5e-3}{\electronvolt/\angstrom}.
For H, spin-polarization was considered.
Other relevant computational conditions were the same as those described in 
% Sec.~\ref{sec:electronic_structures}.
Section~2.2 in the main text.
The zero-point energy (ZPE) of H\textsubscript{2} was evaluated within the harmonic approximation based on the finite displacement method with the displacement of \SI{0.01}{\angstrom}.

\cref{tab:H2} summarizes the such computed \textit{ab initio} bond lengths, ZPEs, and atomization energies, as well as experimental values extracted from literature~\cite{Huber_Book_1979_Molecular,Chase_Book_1998_JANAF,Irikura_JPCRD_2007_Experimental}.
For all the investigated exchange--correlation functionals, the computed values are found to be in reasonable agreement with experiment.
It is therefore safe to use the experimental value of the atomization energy to correct the values for a qualitative comparison in 
% Table~\ref{tab:adsorption_energies}.
Table~1 in the main text.
Since the \textit{ab initio} studies referred to in 
% Table~\ref{tab:adsorption_energies}
Table~1 in the main text
did not account for the contribution of lattice vibrations,
it is fair to correct the values in
% Table~\ref{tab:adsorption_energies}
Table~1 in the main text
with the value \textit{without} ZPE.
We therefore took the experimental atomization energy of H\textsubscript{2} without ZPE, \SI{4.748}{\electronvolt/(H~atom)}, and corrected the values in Refs.~\cite{Mei_JPCC_2011_Density,Yang_JPCC_2012_First,Shetty_JPCC_2017_Computational}
% in Table~\ref{tab:adsorption_energies},
in Table~1 in the main text,
which reference to H\textsubscript{2}
% (Eq.~\eqref{eq:adsorption_energy_H2}),
(eq~2 in the main text),
by \SI{2.374}{\electronvolt/(H~atom)}.

\begin{table}[h]
    \centering
    \caption{Bond lengths (in \si{\angstrom}), ZPEs (in \si{\electronvolt/molecule}), and atomization energies $\Delta_\mathrm{at} E$ (in \si{\electronvolt/molecule}) of {H\textsubscript{2}}.
    Note that PBE-D2 shows essentially the same results as PBE.}
    \label{tab:H2}
    % \begin{ruledtabular}
    \begin{tabular}{cS[table-format=1.3,table-column-width=1.75cm,table-align-text-post=false]
    S[table-auto-round,table-alignment=center,table-format=1.3,table-column-width=1.625cm]
    S[table-auto-round,table-alignment=center,table-format=1.3,table-column-width=1.625cm]
    S[table-auto-round,table-alignment=center,table-format=1.3,table-column-width=1.625cm]}
    \toprule
    &
    {\multirow{2}{*}{Bond length}} &
    {\multirow{2}{*}{ZPE}} &
    \multicolumn{2}{c}{$\Delta_\mathrm{at} E$} \\
    \cmidrule(lr){4-5}
    & & & {w/o ZPE} & {w/ ZPE} \\
    \midrule
    LDA    & 0.766 & 0.258 & 4.90613863 & 4.648 \\
    PBE    & 0.750 & 0.267 & 4.53735896 & 4.270 \\
    PBE-D2 & 0.750 & 0.267 & 4.53738172 & 4.270 \\
\bf SCAN   & 0.741 & 0.275 & 4.66481607 & 4.390 \\
    \midrule
    Exp.   & 0.74144{\footnotemark[1]} & 0.270\footnotemark[2] & 4.748\footnotemark[2]\textsuperscript{,}\footnotemark[3] & 4.478\footnotemark[3] \\
    \bottomrule
    \end{tabular}
    \footnotetext[1]{Ref.~\cite{Huber_Book_1979_Molecular}.}
    \footnotetext[2]{Ref.~\cite{Irikura_JPCRD_2007_Experimental}.}
    \footnotetext[3]{Ref.~\cite{Chase_Book_1998_JANAF}.}
    % \end{ruledtabular}
\end{table}

\clearpage
% \bibliographystyle{apsrev4-2}
% \bibliography{main}
% \bibliography{supplement}
%apsrev4-2.bst 2019-01-14 (MD) hand-edited version of apsrev4-1.bst
%Control: key (0)
%Control: author (8) initials jnrlst
%Control: editor formatted (1) identically to author
%Control: production of article title (0) allowed
%Control: page (0) single
%Control: year (1) truncated
%Control: production of eprint (0) enabled
%